\documentclass[prd,twocolumn,superscriptaddress,showpacs,nofootinbib,amsfonts,amsmath]{revtex4}

\usepackage{graphicx}
\usepackage{dcolumn}
\usepackage{bm}

\begin{document}

\title{Universality of high-energy absorption cross sections for
black holes}

\author{Yves D\'ecanini}
\email{decanini@univ-corse.fr}
\affiliation{\'Equipe Physique Th\'eorique, SPE, UMR 6134 du CNRS
et de l'Universit\'e de Corse,\\
Universit\'e de Corse, Facult\'e des Sciences, BP 52, F-20250
Corte, France}

\author{Gilles Esposito-Far\`ese}
\email{gef@iap.fr}
\affiliation{${\mathcal{G}}{\mathbb{R}}
\varepsilon{\mathbb{C}}{\mathcal{O}}$, Institut d'Astrophysique
de Paris, UMR 7095 du CNRS et de l'Universit\'e Pierre et Marie
Curie-Paris 6, 98bis boulevard Arago, F-75014 Paris, France}

\author{Antoine Folacci}
\email{folacci@univ-corse.fr}
\affiliation{\'Equipe Physique Th\'eorique, SPE, UMR 6134 du CNRS
et de l'Universit\'e de Corse,\\
Universit\'e de Corse, Facult\'e des Sciences, BP 52, F-20250
Corte, France}
\affiliation{Centre de Physique Th\'eorique, UMR 6207 du CNRS et
des Universit\'es Aix-Marseille 1 et 2 et de l'Universit\'e du
Sud Toulon-Var, CNRS-Luminy Case 907, F-13288 Marseille, France}

\date{\today}

\begin{abstract}

We consider the absorption problem for a massless scalar field
propagating in static and spherically symmetric black holes of
arbitrary dimension endowed with a photon sphere. For this wide
class of black holes, we show that the fluctuations of the
high-energy absorption cross section are totally and very simply
described from the properties (dispersion relation and damping) of
the waves trapped near the photon sphere and therefore, in the
eikonal regime, from the characteristics (orbital period and
Lyapunov exponent) of the null unstable geodesics lying on the
photon sphere. This is achieved by using Regge pole techniques. They
permit us to make an elegant and powerful resummation of the
absorption cross section and to extract then all the physical
information encoded in the sum over the partial wave contributions.
Our analysis induces moreover some consequences concerning Hawking
radiation which we briefly report.

\end{abstract}


\maketitle

\section{Introduction}
\label{sec1}

During the last 40 years, the study of absorption of waves and
particles by black holes and by analogous higher-dimensional objects
has received considerable attention, because this topic is directly
relevant to numerous fundamental aspects of classical and quantum
black hole physics which could permit us to progress in our
understanding of spacetime properties. This line of research started
around the 1970s (see
Refs.~\cite{Matzner1968,Mashhoon1973,Starobinsky1973,
StarobinskyChurilov1974,Fabbri1975,Ford1975,
Gibbons1975,Page1976,Unruh1976,Sanchez1978a} for important
pioneering works) motivated by absorption of gravitational waves,
the superradiance phenomenon, and the Hawking effect. In 1997, a
crucial result was obtained by Das, Gibbons and Mathur
\cite{DasGibbonsMathur1997} in connection with string theory: At low
energies, i.e., when the wavelength of the particle is much greater
than the radius of the black hole horizon, the absorption cross
section for a minimally coupled scalar field propagating into a
static and spherically symmetric black hole of arbitrary dimension
presents a universal behavior, in that it reduces to the area of the
black hole horizon. With the extension of this result to the
four-dimensional Kerr-Newman black hole by Maldacena and Strominger
\cite{MaldacenaStrominger1997}, there has been in the last decade an
explosion of the literature concerning this particular theme, with
generalizations for all kinds of fields as well as for various
extensions of general relativity (see, e.g.,
Refs.~\cite{Higuchi2001L,KantiMarchRussell2002,
KantiMarchRussell2003,HarrisKanti2003,JungPark2004,JungPark2005,
DoranEtAL2005,GrainBarrauKanti2005,CrispinoEtAL2007,
HarmakETAL2007,IqbalLiu2009,DolanOliveiraCrispino2009,
CrispinoDolanOliveira2009}, and references therein among hundreds of
articles on the subject), some of these works being done with in
mind the possibility suggested by brane-world theories that the LHC
could behave like a black hole factory (but there are already
significant experimental constraints \cite{CMSCollaboration2010}).

At high energies, it is well known that, in general, the absorption
cross section of a black hole oscillates around a limiting constant
value (see, e.g., Refs.~\cite{Sanchez1978a,
HarrisKanti2003,JungPark2004,JungPark2005,DoranEtAL2005,
GrainBarrauKanti2005,CrispinoEtAL2007,DolanOliveiraCrispino2009,
CrispinoDolanOliveira2009}). The existence of this limiting value
has been understood a long time ago \cite{Darwin1959,Darwin1961,MTW}
in terms of geodesics, and it has also been analyzed for wave
theories \cite{Mashhoon1973,Fabbri1975,Sanchez1978a}: For a black
hole endowed with a photon sphere, the limiting value is exactly the
geometrical cross section of this photon sphere, i.e., the so-called
capture cross section of the black hole. [Here, it is important to
recall that (i)~a photon sphere is a hypersurface on which the
massless particles can orbit the black hole on unstable circular
null geodesics, (ii)~its location corresponds to the local maximum
of the classical effective potential seen by these massless
particles, (iii)~a null geodesic arriving from infinity with the
critical impact parameter $b_c$ approaches the photon sphere
asymptotically by spiralling around it, and (iv)~the geometrical
cross section of the photon sphere is directly related to this
critical impact parameter.] As a consequence, at high energies, the
limiting value of the absorption cross sections for black holes
presents a universal character, but, of course, this does not
suffice to conclude on the universality of the high-energy
absorption cross sections themselves, because the fluctuations
around the limiting value have not yet been explained in terms of
black hole properties.

In fact, it is easy to understand why such a universality result
currently exists at low frequencies and has not yet an equivalent at
high frequencies. In the low-energy regime, the $\ell=0$ partial
wave contribution dominates the absorption cross section (it is the
only non-vanishing contribution) and it is sufficient to solve only
one partial wave equation, in this case, to obtain the universality
result. By contrast, at higher energies, we need to sum over the
full range of partial waves in order to understand the fluctuations
of the absorption cross sections. So it is not possible to provide
immediately a clear physical interpretation of this feature. But in
this paper, by making use of the Regge pole technique (or, in other
words, of the complex angular momentum machinery), we shall realize
a resummation of the absorption cross section and then extract the
physical information encoded in the sum over all the partial wave
contributions. This shall permit us to emphasize the universal
character of the fluctuations at high energies, i.e., to show that
the existence of these fluctuations is a generic feature of black
holes endowed with a photon sphere, which can be described in terms
of the area of the geometrical cross section and the properties of
the waves trapped near the photon sphere. (Here, it is important to
recall that the Regge poles describe the ``surface" or Regge waves
trapped near the photon sphere and that, from their real and
imaginary parts, we obtain, respectively, the nonlinear dispersion
relation of these waves and their damping
\cite{Andersson2,DecaniniFolacci2010a,
DecaniniFolacciRaffaelli2010b}.)

We shall first consider the simple and illuminating example of a
massless scalar field in the Schwarzschild black hole geometry
(Secs.~II and III) and then generalize our analysis to a wide class
of static and spherically symmetric black holes of arbitrary
dimension, which includes various spacetimes of physical interest
such as Schwarzschild-Tangherlini and Reissner-Nordstr\"om black
holes or the canonical acoustic black hole (Sec.~IV). We shall
finally conclude this paper by a short comment concerning the
consequences of our analysis for Hawking radiation, because
absorption and emission phenomena are intimately linked through the
greybody factors of black holes. Throughout this paper, we shall use
units such that $\hbar = c = G = k_B = 1$ and we shall assume a
harmonic time dependence $\exp(-i\omega t)$ for the massless scalar
field.

\section{Absorption cross section of the Schwarzschild black
hole} \label{sec2} The exterior of the Schwarzschild black hole of
mass $M$ can be defined as the manifold ${\cal M} = \hbox{$]-\infty,
+\infty[_t$} \times ]2M,+\infty[_r \times S^2$ with metric $ds^2=
-(1-2M/r)dt^2+ (1-2M/r)^{-1}dr^2+ d\sigma_2^2.$ Here $d\sigma_2^2$
denotes the metric on the unit $2$-sphere $S^2$. We recall that this
black hole has a photon sphere located at $r=3M$ and that the
corresponding critical impact parameter is given by $b_c= 3\sqrt{3}
M$. As a consequence, the geometrical cross section of this black
hole is $\sigma_\mathrm{geo}=\pi b_c^2=27\pi M^2$ (see, e.g.,
Chap.~25 of Ref.~\cite{MTW}).

A massless scalar field $\Phi$ propagating on this gravitational
background satisfies the wave equation $\Box \Phi =0$, which
reduces, after separation of variables and the introduction of
the radial partial wave functions $\phi_{\omega\,\ell}(r)$ with
$\omega >0$ and $\ell=0,1,2, \dots$, to the Regge-Wheeler
equation
\begin{equation}\label{RW}
\frac{d^2 \phi_{\omega \ell}}{dr_\ast^2} + \left[ \omega^2 -
V_{\ell}(r_\ast)\right] \phi_{\omega \ell}=0.
\end{equation}
In Eq.~(\ref{RW}), $r_\ast$ is the so-called tortoise coordinate
defined from the radial Schwarzschild coordinate $r$ by
$dr/dr_\ast=(1-2M/r)$, while $V_{\ell}(r_\ast)$ denotes the
Regge-Wheeler potential given by
\begin{equation}\label{pot_RW_Schw}
V_\ell(r) = \left(\frac{r-2M}{r} \right) \left[
\frac{(\ell+1/2)^2-1/4}{r^2} +\frac{2M}{r^3}\right].
\end{equation}
We recall that $V_\ell(r)$ behaves as a potential barrier with a
maximum located near the photon sphere radius, i.e., near $r=3M$.
For this field and this four-dimensional black hole, the absorption
cross section is given by (see e.g.
Refs.~\cite{Mashhoon1973,Fabbri1975,Sanchez1978a,Andersson1995})
\begin{equation}\label{Sigma_abs}
\sigma_\mathrm{abs}(\omega)=\frac{\pi}{\omega^2}
\sum_{\ell=0}^{+\infty} (2\ell + 1) \Gamma_\ell(\omega),
\end{equation}
where the coefficient $\Gamma_\ell(\omega)$ is the absorption
probability for a particle with energy $\omega$ and angular momentum
$\ell$. It should be also noted that $\Gamma_\ell(\omega)$ plays the
role of a greybody factor when we consider the emission of particles
by the black hole (see also our remarks in the conclusion). When we
only consider the absorption phenomenon, we can define
$\Gamma_\ell(\omega)$ from the solutions $\phi_{\omega \ell}(r)$ of
(\ref{RW}) satisfying the boundary conditions
\begin{equation}\label{BC_Schw}
\phi_{\omega \ell}(r_\ast) \sim \left\{
\begin{array}{cl} T_\ell(\omega) e^{-i\omega r_\ast²}&
\mathrm{for} \ \
r_* \to -\infty,\\
&\\
e^{-i\omega r_\ast} + R_\ell(\omega) e^{+i\omega r_\ast}&
\mathrm{for} \ \ r_* \to +\infty .
\end{array}
\right.
\end{equation}
Here $T_\ell(\omega)$ and $R_\ell(\omega)$ are the transmission
and reflection coefficients for absorption by the Schwarzschild
black hole, and we have more particularly
\begin{equation}\label{Greybodyfactors}
\Gamma_\ell(\omega)= |T_\ell(\omega)|^2.
\end{equation}

In order to extract the information encoded into the sum over all
the partial wave contributions (\ref{Sigma_abs}), we first transform
the angular momentum $\ell$ into a complex number $\lambda= \ell+
1/2$ and we construct the analytic extension
$\Gamma_{\lambda-1/2}(\omega)$ of the greybody factor
$\Gamma_\ell(\omega)$. It should be noted that there is no unique
way to achieve this construction. Such a problem is well-known by
practitioners of complex angular momentum techniques (see, e.g.,
Chap.~13 of Ref.~\cite{New82}), and we recall that, in general, a
suitable interpolation is provided by the simplest extension and is
ultimately justified by the results it provides. In our case, we
have adopted the following prescription:
\begin{equation}\label{Greybodyfactors_interpolation}
\Gamma_{\lambda-1/2}(\omega)=T_{\lambda-1/2}(\omega) {\overline
{T_{{\overline \lambda}-1/2}(\omega)}},
\end{equation}
where $T_{\lambda-1/2}(\omega)$ is the transmission coefficient for
the problem defined by (\ref{RW}), (\ref{pot_RW_Schw}) and
(\ref{BC_Schw}) with $\ell \to \lambda- 1/2$ and where ${\overline
{T_{{\overline \lambda}-1/2}(\omega)}}$ is the transmission
coefficient for the same problem but with $\phi_\ell(\omega) \to
{\overline {{\phi}_\ell(\omega)}}$ and $\ell \to {\overline
\lambda}-\frac{1}{2}$ (here the bar denotes complex conjugation).
This prescription allows us to work with the same Regge-Wheeler
equation in both cases and gives a consistent analytic extension
because $\Gamma_{\lambda-1/2}$ is then clearly a function of
$\lambda$ but not $\overline \lambda$. On the contrary, note that
$|T_\ell(\omega)|^2$ could not be extended as
$T_{\lambda-1/2}(\omega) {\overline {T_{\lambda-1/2}(\omega)}}$,
because it would be real on all the complex plane and thereby
non-analytic since it is not constant. Our prescription will also be
justified by another argument below Eq.~(\ref{PRapproxWKB}).

It is now interesting to emphasize some important and useful
properties of the function $\Gamma_{\lambda-1/2}(\omega)$:

(i)~$\Gamma_{\lambda-1/2}(\omega)$ is an even function of $\lambda$.
Indeed, $\phi_{\lambda-1/2}(\omega)$ and
$\phi_{-\lambda-1/2}(\omega)$ are both solutions of the same
Regge-Wheeler problem (\ref{RW})-(\ref{pot_RW_Schw}) and satisfy the
same boundary conditions (\ref{BC_Schw}). As a consequence,
$T_{\lambda-1/2}(\omega)$ and $T_{-\lambda-1/2}(\omega)$ can only
differ by a phase factor. This phase factor is necessarily of the
type $\exp[i \theta(\lambda)]$, where $\theta(\lambda)$ is a real
function. From the prescription
(\ref{Greybodyfactors_interpolation}), we then immediately obtain
$\Gamma_{-\lambda-1/2}(\omega)=\Gamma_{\lambda-1/2}(\omega)$.

(ii)~The singularities (simple poles) of
$\Gamma_{\lambda-1/2}(\omega)$ are symmetrically distributed with
respect to the real $\lambda$ axis [as well as symmetrically
distributed with respect to the origin $O$ of the complex $\lambda$
plane due to property (i)]: Those associated with
$T_{\lambda-1/2}(\omega)$ lie in the first (and in the third)
quadrant of the complex $\lambda$ plane and are also the Regge poles
$\lambda_n(\omega)$, with $n \in \mathbb{N}^*$ (where $\mathbb{N}^*
\equiv \mathbb{N}\setminus\lbrace{0\rbrace}$), of the $S$ matrix of
the Schwarzschild black hole (see Refs.~\cite{Andersson1,Andersson2,
DecaniniFJ_cam_bh,DecaniniFolacci2010a}, and more particularly
Fig.~1 of Ref.~\cite{DecaniniFolacci2010a}). Their complex
conjugates ${\overline {\lambda_n(\omega)}}$ are associated with
${\overline {T_{{\overline \lambda}-1/2}(\omega)}}$ and they lie in
the fourth (and in the second) quadrant of the complex $\lambda$
plane.

(iii)~The residues of $\Gamma_{\lambda-1/2}(\omega)$ at the poles
$\lambda_n(\omega)$ and ${\overline {\lambda_n(\omega)}}$ are
complex conjugates of each other and we have in particular
\begin{eqnarray}\label{Sigma_abs_PR_b}
&&\gamma_n(\omega) \equiv {\mathrm{residue}[
\Gamma_{\lambda-1/2}(\omega)]}_{\lambda=\lambda_n(\omega)}
\nonumber\\
&&\phantom{\gamma_n(\omega)}={\mathrm{residue}[
T_{\lambda-1/2}(\omega)]}_{\lambda=\lambda_n(\omega)}{\overline {
T_{{\overline {\lambda_n(\omega)}-1/2}}(\omega)}}.\qquad
\end{eqnarray}

By means of the usual ``half-range" Poisson sum formula
\begin{equation}\label{HR_Poisson}
\sum_{\ell=0}^{+\infty} F(\ell+1/2)=\sum_{m=-\infty}^{+\infty}
(-1)^m \int_0^{+\infty} F(\lambda)e^{i 2 \pi m \lambda}\,d\lambda,
\end{equation}
we can rewrite the sum (\ref{Sigma_abs}) as
\begin{eqnarray}\label{sigma_abs_HR_Poisson}
&&\sigma_\mathrm{abs}(\omega)=\frac{2\pi}{\omega^2}
\int_0^{+\infty} \lambda \Gamma_{\lambda-1/2}(\omega)
\,d\lambda \nonumber\\
&&\qquad + \frac{2\pi}{\omega^2}\sum_{m=1}^{+\infty}
\int_0^{+\infty}
\lambda \Gamma_{\lambda-1/2}(\omega)e^{i 2m \pi (\lambda-1/2)}
\,d\lambda \nonumber\\
&&\qquad + \frac{2\pi}{\omega^2}\sum_{m=1}^{+\infty}
\int_0^{+\infty} \lambda \Gamma_{\lambda-1/2}(\omega)e^{-i 2m \pi
(\lambda-1/2)}\,d\lambda.\qquad
\end{eqnarray}
The integrals in the second term of (\ref{sigma_abs_HR_Poisson}) can
be evaluated by using Cauchy's theorem. To do so, we close the path
along the positive real axis with a quarter circle at infinity in
the first quadrant of the complex angular momentum plane and a path
along the positive imaginary axis going from $+i\infty$ to $0$. The
integrals in the third term of (\ref{sigma_abs_HR_Poisson}) can be
evaluated similarly but now by closing the path along the positive
real axis in the fourth quadrant of the complex angular momentum
plane. By noting that all the contributions of the contours at
infinity vanish and by taking into account the singularities of the
integrands lying in the first and fourth quadrants of the complex
$\lambda$ plane, i.e., the poles of the greybody factor
$\Gamma_{\lambda-1/2}(\omega)$, we obtain
\begin{eqnarray}\label{sigma_abs_SC1}
&&\sigma_\mathrm{abs}(\omega)=\frac{2\pi}{\omega^2}
\int_0^{+\infty} \lambda \Gamma_{\lambda-1/2}(\omega)
\,d\lambda \nonumber\\
&&\quad +\frac{8\pi^2}{\omega^2}\,\mathrm{Re} \left(
\sum_{m=1}^{+\infty} \sum_{n=1}^{+\infty} i\,\lambda_n(\omega)
\gamma_n(\omega)\, e^{i2m \pi[\lambda_n(\omega)-1/2]} \right)
\nonumber\\
&&\quad + \frac{2\pi}{\omega^2} \sum_{m=1}^{+\infty} \left(
\int_0^{+i\infty} \lambda \Gamma_{\lambda-1/2}(\omega)\, e^{i2m
\pi[\lambda -1/2]}\,d\lambda \right. \nonumber\\
&&\qquad \qquad \left. + \int_0^{-i\infty} \lambda
\Gamma_{\lambda-1/2}(\omega)\, e^{- i2m \pi[\lambda -1/2]}
\,d\lambda \right).\qquad
\end{eqnarray}
In Eq.~(\ref{sigma_abs_SC1}), the real part symbol appears
because the singularities of $\Gamma_{\lambda-1/2}(\omega)$ lie
symmetrically distributed with respect to the real $\lambda$ axis
[see properties (ii) and (iii) mentioned in the previous
paragraph].

It should be noted that we can simplify (\ref{sigma_abs_SC1}) by
using the relations (for $a \in {\mathbb {R}}$)
\begin{subequations}\label{devS sin}
\begin{eqnarray}
&&\sum_{m=1}^{+\infty} e^{i 2m \pi (z -a)}=\frac{i}{2}\frac{e^{i
\pi (z-a)}}{\sin [\pi (z-a)] } \nonumber\\
&&\qquad\qquad\qquad\qquad\qquad \quad \mathrm{valid\, if}\,
\mathrm{Im} \ z > 0, \label{devS sin_a}\quad\\
&&\sum_{m=1}^{+\infty} e^{-i 2m \pi (z -a)}=
-\frac{i}{2}\frac{e^{-i
\pi (z-a)}}{\sin [\pi (z-a)] } \nonumber\\
&&\qquad\qquad\qquad\qquad\qquad \quad \mathrm{valid\, if}\,
\mathrm{Im} \ z < 0, \label{devS sin_b}\quad
\end{eqnarray}
\end{subequations}
as well as the parity of $\Gamma_{\lambda-1/2}(\omega)$. We then
obtain
\begin{eqnarray}\label{sigma_abs_SC2}
\sigma_\mathrm{abs}(\omega)&=&\frac{2\pi}{\omega^2}
\int_0^{+\infty} \lambda \Gamma_{\lambda-1/2}(\omega)
\,d\lambda \nonumber\\
&&-\frac{4\pi^2}{\omega^2}\,\mathrm{Re} \left(
\sum_{n=1}^{+\infty} \frac{ e^{i\pi[\lambda_n(\omega)-1/2]}\,
\lambda_n(\omega) \gamma_n(\omega)
}{\sin[\pi (\lambda_n(\omega)-1/2)]} \right)\nonumber\\
&&- \frac{2\pi}{\omega^2} \int_0^{+i\infty} \frac{e^{i
\pi \lambda}}{\cos (\pi \lambda )}\lambda
\Gamma_{\lambda-1/2}(\omega)\,d\lambda.
\end{eqnarray}
It is important to note that (\ref{sigma_abs_SC2}) [or
equivalently (\ref{sigma_abs_SC1})] is an exact expression for
the total absorption cross section of the Schwarzschild black
hole. Indeed, until now, we have not made any approximation. We
have just expressed (\ref{Sigma_abs}) in a different way. It is
also worth pointing out, for the mathematically inclined reader,
that (\ref{sigma_abs_SC1}) and (\ref{sigma_abs_SC2}) are
reminiscent of trace formulas considered in semiclassical
analysis and in analytic number theory.

The residue series over the Regge poles of the greybody factor
matrix appearing in (\ref{sigma_abs_SC2}), and given by
\begin{equation}\label{Sigma_abs_PR_a}
\sigma_\mathrm{abs}^\mathrm{RP}(\omega)=-\frac{4\pi^2}{\omega^2}\,
\mathrm{Re} \left( \sum_{n=1}^{+\infty} \frac{
e^{i\pi[\lambda_n(\omega)-1/2]}\,\lambda_n(\omega)
\gamma_n(\omega)}{\sin[\pi (\lambda_n(\omega)-1/2)]} \right),
\end{equation}
provides the oscillating part of the total absorption cross
section. From the first and third terms of (\ref{sigma_abs_SC2}),
we can extract the geometrical cross section of the black hole
$\sigma_\mathrm{geo}=27\pi M^2$ (the constant limit of the
absorption cross section, which therefore does not play any role
in the fluctuations), together with a negligible
$\mathcal{O}(1/\omega^2)$ contribution at high frequencies.
Indeed, on the positive real $\lambda$ axis, we can consider that
the greybody factor (the absorption probability) is roughly given
by
\begin{equation}\label{GBF_DeWitt}
\Gamma_{\lambda-1/2}(\omega) =
\Theta (3\sqrt{3}M \omega - \lambda),
\end{equation}
where $\Theta$ is the Heaviside step function, because we know that
partial waves characterized by an energy $\omega$ and an angular
momentum $\ell$ are totally reflected if $\ell +1/2 \gg
3\sqrt{3}M\omega$ and totally absorbed if $\ell +1/2 \ll
3\sqrt{3}M\omega$. Then, by inserting (\ref{GBF_DeWitt}) into the
first term of (\ref{sigma_abs_SC2}) we obtain exactly the
geometrical cross section of the black hole. Of course, it is
possible to consider more elaborate models for the greybody factor.
For example [see Eq.~(\ref{Modules Tell}) below], we could assume
that, for $\lambda \in [0,+\infty[$,
\begin{equation}\label{GBF_Gaina}
\Gamma_{\lambda-1/2}(\omega)= \frac{1}{1+\exp\left[-2\pi
\left(\frac{27M^2\omega^2-\lambda^2}{2\lambda} \right) \right]}.
\end{equation}
This expression reduces to (\ref{GBF_DeWitt}) for low and high
frequencies and it moreover provides an accurate description of
the transition between these two regimes, i.e., for $\lambda
\approx 3\sqrt{3}M\omega$. By inserting (\ref{GBF_Gaina}) into
the first term of (\ref{sigma_abs_SC2}) we obtain again the
geometrical cross section of the black hole but, now, with in
addition a correction given by $(\pi/6)/\omega^2$. This kind of
result seems to be very robust, i.e., independent of any realist
model for the greybody factor, and we shall consider that we have
always for the first term of (\ref{sigma_abs_SC2})
\begin{equation}\label{FristTerm}
\frac{2\pi}{\omega^2} \int_0^{+\infty} \lambda
\Gamma_{\lambda-1/2}(\omega)\,d\lambda= 27\pi M^2 +
\mathcal{O}(1/\omega^2).
\end{equation}
Let us also consider the third term of (\ref{sigma_abs_SC2}).
$\Gamma_{\lambda-1/2}(\omega)$ is bounded on the positive imaginary
$\lambda$ axis (it has no poles on this axis, is equal to 1 for
$\lambda=0$ and vanishes at $+ i\infty$). In fact, we have checked
numerically [but a mathematical proof involving the fact that
$1/\Gamma_{\lambda-1/2}(\omega)$ is an entire function of $\lambda$
could be also provided] that $\forall \omega >0$ and $\forall
-i\lambda \in [0,+ \infty[$, $0< \Gamma_{\lambda-1/2}(\omega) \le
1$. As a consequence, we have
\begin{equation}\label{ThirdTerm}
0< - \frac{2\pi}{\omega^2} \int_0^{+i\infty} \frac{e^{i \pi
\lambda}}{\cos (\pi \lambda )}\lambda \Gamma_{\lambda-1/2}(\omega)
\,d\lambda < \frac{(\pi /12)}{\omega^2}.
\end{equation}
In fact, one can even go further and note that the function $\lambda
e^{i \pi \lambda} /\cos (\pi \lambda )$ has only significant values
for $-i \lambda <1 $ or, in other words, when
$\Gamma_{\lambda-1/2}(\omega) \approx 1$. As a consequence, we could
consider that
\begin{equation}\label{ThirdTerm_bis}
- \frac{2\pi}{\omega^2} \int_0^{+i\infty} \frac{e^{i \pi
\lambda}}{\cos (\pi \lambda )}\lambda \Gamma_{\lambda-1/2}(\omega)
\,d\lambda \approx \frac{(\pi /12)}{\omega^2}. \nonumber
\end{equation}
{}From (\ref{FristTerm}) and (\ref{ThirdTerm}), we can then
replace (\ref{sigma_abs_SC1}) by
\begin{eqnarray}\label{sigma_abs_SC3}
\sigma_\mathrm{abs}(\omega)&=&27\pi M^2 \nonumber\\
&&-\frac{4\pi^2}{\omega^2}\,\mathrm{Re} \left(
\sum_{n=1}^{+\infty} \frac{ e^{i\pi[\lambda_n(\omega)-1/2]}\,
\lambda_n(\omega) \gamma_n(\omega)}
{\sin[\pi (\lambda_n(\omega)-1/2)]} \right)\nonumber\\
&&+ \mathcal{O}(1/\omega^2).
\end{eqnarray}
Of course, it would be useful to also capture analytically the last
term $\mathcal{O}(1/\omega^2)$, because it can play a numerically
significant role at low frequencies (see, e.g.,
Fig.~\ref{fig:AcousticBH} below), although it is negligible at high
enough frequencies. However, it depends in a subtle way on the exact
form of $\Gamma_{\lambda-1/2}(\omega)$ in the complex $\lambda$
plane and therefore on the specific black hole under consideration.
We shall not compute it in the present paper.

\begin{figure}
\includegraphics[width=8.5cm]{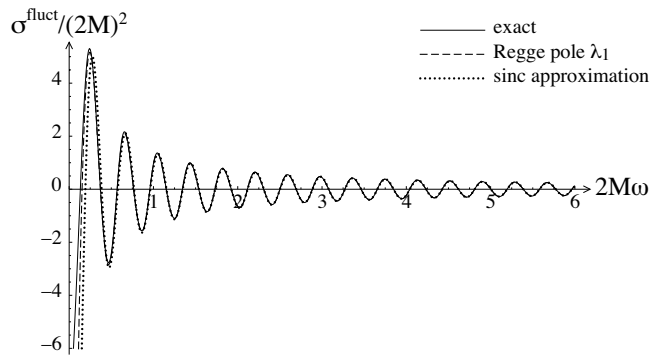}
\caption{\label{fig:SecAbsExactAndPR1}Fluctuations of the total
absorption cross section, $\sigma^\text{fluct} \equiv
\sigma_\text{abs} - \sigma_\text{geo}$, for a massless scalar
field propagating in the Schwarzschild geometry.}
\end{figure}

\begin{figure}
\includegraphics[width=8.5cm]{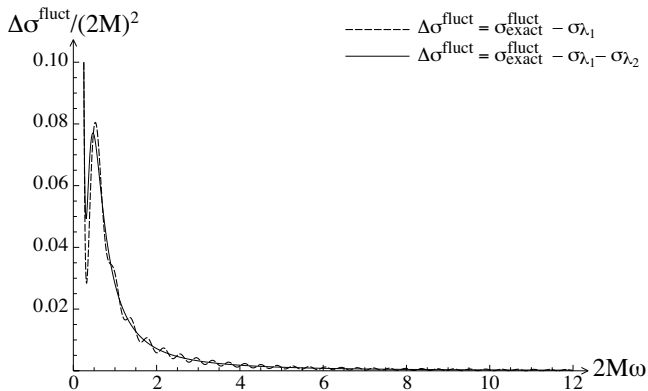}
\caption{\label{fig:DiffSigma}Differences between the exact
fluctuations of the total absorption cross section, for a
massless scalar field propagating in the Schwarzschild geometry,
and the contributions of the first and the first two Regge
poles.}
\end{figure}

In Fig.~\ref{fig:SecAbsExactAndPR1} are displayed the fluctuations
of the total absorption cross section, obtained by subtracting the
geometrical cross section of the Schwarzschild black hole,
$\sigma_\mathrm{geo}=27\pi M^2$, from the full sum over the partial
waves~(\ref{Sigma_abs}). We also display the contribution of the
first Regge pole corresponding to the term $n=1$ in
Eq.~(\ref{Sigma_abs_PR_a}). Both cross sections were obtained by
solving numerically the problem defined by (\ref{RW}),
(\ref{pot_RW_Schw}) and (\ref{BC_Schw}) for integer and complex
values of the angular momentum. The agreement of the two graphs is
remarkable, and we could already consider that the fluctuations of
the Schwarzschild absorption cross section around the geometrical
cross section are very well described by the first Regge pole
contribution. This is further illustrated in
Fig.~\ref{fig:DiffSigma}, which shows that the difference of the two
curves is numerically small with respect to their amplitude in
Fig.~\ref{fig:SecAbsExactAndPR1}. We also observe that small
oscillations remain present in Fig.~\ref{fig:DiffSigma}, but they
are eliminated by taking into account the contributions of the other
Regge poles in Eq.~(\ref{Sigma_abs_PR_a}). Actually, the
contribution of the $n=2$ term is sufficient to explain the effects
displayed in Fig.~\ref{fig:DiffSigma}. The remaining smooth
contribution to $\Delta\sigma^\text{fluct}/(2M)^2$ comes from the
$\mathcal{O}(1/\omega^2)$ term discussed previously. For high
frequencies, it behaves as $\pi/(12 \sqrt{3} M \omega)^2$.

\section{High-energy analytic formula
} \label{sec3} To end with the Schwarzschild black hole, we now
derive a simple formula which describes numerically and physically
the fluctuations of the high-energy absorption cross section.
Formally, it is valid in the eikonal regime (i.e., for very high
frequencies), but we can actually use it even for rather low
frequencies. We first recall that the Regge poles of the $S$ matrix,
as the associated quasinormal complex frequencies, are due to
tunneling near the top of the potential barrier described by
(\ref{pot_RW_Schw}) (see Refs.~\cite{DecaniniFolacci2010a,
SchutzWill,Iyer1,Iyer2}). As a consequence, even if it has been
obtained for $\omega >0$ and $\ell \in {\mathbb {N}}$
\cite{Iyer1,Iyer2}, we can start from the formula
\begin{equation}\label{Modules Tell}
\Gamma_\ell(\omega)=|T_\ell(\omega)|^2=\frac{1}{1+\exp\left[-2\pi
\frac{\omega^2-V_0(\ell)}{\sqrt{-2V^{''}_0(\ell)}}\right]}\,,
\end{equation}
where
\begin{subequations}\label{V_0 et dersec}
\begin{eqnarray}
V_0(\ell)&\equiv&\left.
V_{\ell}(r_*)\right|_{r_*={(r_*)}_0}=
\frac{(\ell+1/2)^2}{27M^2} + \underset{\ell \to
+\infty}{\cal O}(1),\nonumber\\
&&\\
V^{''}_0(\ell)&\equiv&\left. \frac{d^2}{{dr_*}^2 }
V_{\ell}(r_*)\right|_{r_*={(r_*)}_0}
\nonumber\\
&=&-\frac{2(\ell+1/2)^2}{(27M^2)^2} +
\underset{\ell \to +\infty}{\cal O}(1),
\end{eqnarray}
\end{subequations}
which provides the absorption probability for a particle when
$\omega^2 \approx V_0(\ell)$. Here ${(r_*)}_0$ denotes the
maximum of the function $V_{\ell}(r_*)$ obtained from
(\ref{pot_RW_Schw}) by using the tortoise coordinate. We now
extend (\ref{Modules Tell}) to complex angular momenta. It is
then easy to show that the singularities of
$\Gamma_{\lambda-1/2}(\omega)$ are the Regge poles
$\lambda_n(\omega)$, with $n \in \mathbb{N}^*$, of the
Schwarzschild black hole $S$ matrix, given by the approximation
\cite{DecaniniFJ_cam_bh,DecaniniFolacci2010a}
\begin{equation}\label{PRapproxWKB}
\lambda_n(\omega) = 3\sqrt{3}M\omega
+ i \left(n-\frac{1}{2}\right)
+ \underset{2M\omega \to +\infty}{\cal
O}\left(\frac{1}{2M\omega}\right),
\end{equation}
and lying in the first quadrant of the complex $\lambda$ plane,
as well as their complex conjugates ${\overline
\lambda}_n(\omega)$ [and obviously also the opposites of both,
$-\lambda_n(\omega)$ and $-{\overline \lambda}_n(\omega)$, but
they do not contribute to our calculation
(\ref{sigma_abs_HR_Poisson})--(\ref{sigma_abs_SC3})]. This result
is in agreement with those obtained above and validates the
prescription used in order to construct the analytic extension
$\Gamma_{\lambda-1/2}(\omega)$ of the greybody factor
$\Gamma_\ell(\omega)$. Furthermore, we have immediately for the
residue (\ref{Sigma_abs_PR_b})
\begin{equation}\label{Sigma_abs_PR_b_bis}
\gamma_n(\omega) = -\frac{1}{2\pi} + \underset{2M\omega \to
+\infty}{\cal O}\left(\frac{1}{2M\omega}\right).
\end{equation}
Then, by inserting (\ref{PRapproxWKB}) and
(\ref{Sigma_abs_PR_b_bis}) into (\ref{Sigma_abs_PR_a}), using
(\ref{devS sin_a}) and keeping only the contribution of the first
Regge pole, we obtain for the high-energy behavior of the
oscillating part of the absorption cross section the very simple
formula
\begin{equation}\label{Sigma_abs_PR_a_APP}
\sigma_\mathrm{abs}^\mathrm{osc}(\omega)=-8 \pi\, e^{-\pi}\,
\sigma_\mathrm{geo} \times \text{sinc}\left[2\pi (3\sqrt{3}
M)\omega\right],
\end{equation}
where $\text{sinc}\, x \equiv (\sin x)/x$ is the sine cardinal. The
constant factor multiplying this sinc function involves the
geometrical cross section of the black hole
$\sigma_\mathrm{geo}=27\pi M^2$ and the argument of the sinc
involves the orbital period, $2\pi (3\sqrt{3} M)$, of a massless
particle orbiting the black hole on the photon sphere (see, e.g.,
Refs.~\cite{DecaniniFJ_cam_bh,DecaniniFolacci2010a,
CardosoMirandaBertietal2009}). In fact, the coefficient $8 \pi
e^{-\pi}$ involves the Lyapunov exponent of the geodesic followed by
the particle, but this does not appear clearly here [see our
discussion below Eq.~(\ref{Sigma_abs_PR_a_APP_dimD}) for more
precision]. In Fig.~\ref{fig:SecAbsExactAndPR1}, we have also
displayed the eikonal cross section (\ref{Sigma_abs_PR_a_APP}). The
agreement with the exact result is very good, even for low
frequencies. Of course, a more detailed numerical study shows that
(\ref{Sigma_abs_PR_a_APP}) is actually much less accurate than
(\ref{Sigma_abs_PR_a}), but it is anyway very nice to have such a
simple formula to describe the oscillations of the absorption cross
section of the Schwarzschild black hole. Moreover, this formula
permits us to interpret naturally the period of the maxima (or the
minima, or the zeros) of the fluctuations in terms of constructive
interferences of the ``surface waves" trapped near the photon sphere
(see also Refs.~\cite{DecaniniFJ_cam_bh} and
\cite{DecaniniFolacci2010a} for related aspects). It is interesting
to note that, a long time ago, S\'anchez had obtained a fit of the
numerical absorption cross section of the Schwarzschild black hole
which involved the sinc function of Eq.~(\ref{Sigma_abs_PR_a_APP})
\cite{Sanchez1978a}. Her result was obtained from purely numerical
considerations. Here, we have behind formula
(\ref{Sigma_abs_PR_a_APP}) a powerful and solid theory with a clear
physical interpretation, and it can easily be generalized as we
shall now see.

\section{Absorption cross sections of arbitrary static and
spherically symmetric black holes}
\label{sec4}

We now consider a static and spherically symmetric black hole of
arbitrary dimension $d \ge 4$, endowed with a photon sphere. It is
defined as the manifold ${\cal M} = \hbox{$]-\infty, +\infty[_t$}
\times ]r_h,+\infty[_r \times S^{d-2}$ with metric $ds^2= -f(r)dt^2+
f(r)^{-1}dr^2+ d\sigma_{d-2}^2$, where $d\sigma_{d-2}^2$ denotes the
metric on the unit $(d-2)$-sphere $S^{d-2}$. We furthermore assume
that $r_h$ is a simple root of $f(r)$ and that we have $f(r)>0$ for
$r > r_h$ and $\underset{r \to +\infty}{\lim}f(r)=1$. We also assume
that there exists a value $r_c \in \hbox{$]r_h,+\infty[$}$ for which
$f'(r_c) - \frac{2}{r_c}f(r_c)=0$ and
$f''(r_c)-\frac{2}{r_c^2}f(r_c) <0$. In other words, the spacetime
considered is an asymptotically flat black hole with an event
horizon at $r_h$, its exterior corresponding to $r > r_h $, and the
assumptions on $r_c$ imply the existence of a photon sphere located
at $r_c$ which is the support of unstable circular null geodesics
(see Ref.~\cite{DecaniniFolacciRaffaelli2010b} for more details on
these different assumptions). It should also be noted that now the
critical impact parameter is given by $b_c= r_c/\sqrt{f(r_c)}$ and
that the geometrical cross section of this black hole is
$\sigma_\mathrm{geo}=\pi^{(d-2)/2} b_c^{d-2}/\Gamma(d/2)$ (see,
e.g., Ref.~\cite{HarrisKanti2003}).

A massless scalar field $\Phi$ propagating on this gravitational
background satisfies the wave equation $\Box \Phi =0$, which
reduces, after separation of variables and the introduction of
the radial partial wave functions $\phi_{\omega\,\ell}(r)$ with
$\omega >0$ and $\ell=0,1,2, \dots$, to the Regge-Wheeler
equation (\ref{RW}), but instead of (\ref{BC_Schw}) we take
\begin{eqnarray}\label{EffectivePot_dimD}
V_\ell(r)&=&f(r) \left[ \frac{[\ell+(d-3)/2]^2-[(d-3)/2]^2}{r^2}
\right. \nonumber\\
&&+ \left. \frac{(d-2)(d-4)}{4r^2}f(r) +
\left(\frac{d-2}{2r}\right)f'(r) \right],\qquad
\end{eqnarray}
and the tortoise coordinate $r_\ast$ is now defined from $r$ by
the relation $dr_\ast/dr=1 / f(r)$.

{}From Eq.~(9) of Ref.~\cite{Gubser1997}, we can write the
corresponding $d$-dimensional absorption cross section of the
black hole in the form
\begin{eqnarray}\label{Sigma_abs_dimD}
\sigma_\mathrm{abs}(\omega)&=&\frac{\pi^{(d-2)/2}}{\Gamma
\left[(d-2)/2\right]
\omega^{d-2}}
\nonumber\\
&&\times \sum_{\ell=0}^{+\infty} \frac{(\ell+d-4)!}{\ell!}
\left(2\ell + d-3\right) \Gamma_\ell(\omega),\quad
\end{eqnarray}
where $\Gamma_\ell(\omega)$ for $\ell = 0, 1, \dots $
are the greybody factors defined now by (\ref{RW}),
(\ref{EffectivePot_dimD}), (\ref{BC_Schw}) and
(\ref{Greybodyfactors}). [The Euler $\Gamma$ function should not
be confused with these greybody factors $\Gamma_\ell$, bearing a
lower index.]

Various equivalent versions of (\ref{Sigma_abs_dimD}) can be
obtained by transforming the angular momentum $\ell$ into a
complex number $\lambda = \ell + (d-3)/2$ and by constructing the
analytic extension $\Gamma_{\lambda -(d-3)/2}(\omega)$ of
$\Gamma_\ell (\omega)$. In the $d$-dimensional context, we shall
consider that
\begin{equation}\label{Greybodyfactors_interpolation_d_dim}
\Gamma_{\lambda-(d-3)/2}(\omega)=T_{\lambda-(d-3)/2}(\omega)
{\overline {T_{{\overline \lambda}-(d-3)/2}(\omega)}},
\end{equation}
where $T_{\lambda-(d-3)/2}(\omega)$ is the transmission
coefficient for the problem defined by (\ref{RW}),
(\ref{EffectivePot_dimD}) and (\ref{BC_Schw}) with $\ell \to
\lambda- (d-3)/2$. {\it Mutatis mutandis}, the properties of
$\Gamma_{\lambda-(d-3)/2}(\omega)$ are identical to those of
$\Gamma_{\lambda-1/2}(\omega)$ emphasized in Sec.~II. In
particular:

(i)~$\Gamma_{\lambda-(d-3)/2}(\omega)$ is an even function of
$\lambda$.

(ii)~The singularities of $\Gamma_{\lambda-(d-3)/2}(\omega)$ are
symmetrically distributed with respect to the real $\lambda$ axis
as well as symmetrically distributed with respect to the origin
$O$ of the complex $\lambda$ plane. Furthermore, if we denote as
$\lambda_n(\omega)$, with $n \in \mathbb{N}^*$, those lying in
the first quadrant of the complex $\lambda$ plane (the so-called
Regge poles), those lying in the fourth quadrant are their
complex conjugates ${\overline {\lambda_n(\omega)}}$, with $n \in
\mathbb{N}^*$.

(iii)~The residues of $\Gamma_{\lambda-(d-3)/2}(\omega)$ at the
poles $\lambda_n(\omega)$ and ${\overline {\lambda_n(\omega)}}$
are complex conjugates of each other, and we shall denote
\begin{equation}\label{Sigma_abs_PR_b_ddim}
\gamma_n(\omega) \equiv {\mathrm{residue}[
\Gamma_{\lambda-(d-3)/2}(\omega)]}_{\lambda=\lambda_n(\omega)}.
\end{equation}

In the $d$-dimensional context, it is not possible to sum the
series (\ref{Sigma_abs_dimD}) by using one of the usual Poisson
sum formula. However, we shall succeed in generalizing
(\ref{sigma_abs_HR_Poisson}) by using a modified version of the
Sommerfeld-Watson transformation \cite{New82}. Indeed, we can
write
\begin{eqnarray}\label{Sigma_abs_SW_dimD}
\sigma_\mathrm{abs}(\omega)&=&\frac{i\,
\pi^{(d-2)/2}}{\Gamma\left[(d-2)/2\right] \omega^{d-2}}
\int_{\cal C} \frac{\Gamma[\lambda+(d-3)/2]}{\Gamma
[\lambda-(d-5)/2]}\nonumber\\
&&\times \frac{e^{i\pi [\lambda -(d-3)/2]} }{\sin [\pi
(\lambda-(d-3)/2)]}\,\lambda \Gamma_{\lambda-(d-3)/2}(\omega)\,
d\lambda, \nonumber\\
&&
\end{eqnarray}
where ${\cal C}=]+\infty -i\epsilon,-i\epsilon] \cup
[-i\epsilon,+i\epsilon] \cup [+i\epsilon, +\infty +i\epsilon[$
(here, we have $\epsilon \to 0_+$). We can recover
(\ref{Sigma_abs_dimD}) from (\ref{Sigma_abs_SW_dimD}) by using
Cauchy's theorem and by noting that the poles of the integrand in
(\ref{Sigma_abs_SW_dimD}) are the semi-integers (if $d$ is even)
or the integers (if $d$ is odd) $\lambda = \ell + (d-3)/2$, with
$\ell \in \mathbb{N}$. Here, it is important to remark that the
poles of the function $1/\sin [\pi (\lambda-(d-3)/2)]$ into the
interval $[0,(d-3)/2[$ do not contribute because they are
neutralized by the zeros of the function $\Gamma[\lambda+(d-3)/2]
/ \Gamma [\lambda-(d-5)/2]$, which can be written in the form
\begin{equation}\label{Rapp_GsurG}
\frac{\Gamma[\lambda+(d-3)/2]}{\Gamma [\lambda-(d-5)/2]} =\left\{
\begin{array}{cl}&\prod_{k=1}^{(d-3)/2-1/2}\,
[\lambda^2 - (k-1/2)^2]
\\
&\qquad \mathrm{for}\,\,
d\,\,\mathrm{even\,\, and}\,\,\ge 4,\\
&\\
&\lambda\,\prod_{k=1}^{(d-3)/2-1}\, [\lambda^2 - k^2]\\
&\qquad \mathrm{for}\,\, d\,\,\mathrm{odd\,\, and}
\,\,\ge 5 .
\end{array}
\right.
\end{equation}
On the part $[+i\epsilon, +\infty +i\epsilon[$ of the integration
contour ${\cal C}$ where $\mathrm{Im} \ \lambda > 0$, we can use
(\ref{devS sin_a}). On the part $]+\infty -i\epsilon,$
$-i\epsilon]$ of this integration contour where $\mathrm{Im} \
\lambda < 0$, we can use (\ref{devS sin_b}) in the form
\begin{equation} \label{devS sin_b_ddim}
\frac{e^{i
\pi [z-(d-3)/2]}}{\sin [\pi (z-(d-3)/2)] }=2i
+ 2i\sum_{m=1}^{+\infty} e^{-i 2m \pi [z -(d-3)/2]}.
\end{equation}
We can then replace (\ref{Sigma_abs_SW_dimD}) by
\begin{widetext}
\begin{eqnarray}\label{sigma_abs_SC0_ddim}
\sigma_\mathrm{abs}(\omega)&=&\frac{2
\pi^{(d-2)/2}}{\Gamma\left[(d-2)/2\right] \omega^{d-2}}
\int_0^{+\infty} \frac{\Gamma[\lambda+(d-3)/2]}{\Gamma
[\lambda-(d-5)/2]}\,\lambda \Gamma_{\lambda-(d-3)/2}(\omega)
\,d\lambda \nonumber\\
&&+ \frac{2 \pi^{(d-2)/2}}{\Gamma\left[(d-2)/2\right]
\omega^{d-2}}\sum_{m=1}^{+\infty} \int_0^{+\infty}
\frac{\Gamma[\lambda+(d-3)/2]}{\Gamma
[\lambda-(d-5)/2]}\,\lambda \Gamma_{\lambda-(d-3)/2}(\omega)
e^{i 2m \pi [\lambda-(d-3)/2]}\,d\lambda \nonumber\\
&&+ \frac{2 \pi^{(d-2)/2}}{\Gamma\left[(d-2)/2\right]
\omega^{d-2}}\sum_{m=1}^{+\infty} \int_0^{+\infty}
\frac{\Gamma[\lambda+(d-3)/2]}{\Gamma [\lambda-(d-5)/2]}\,\lambda
\Gamma_{\lambda-(d-3)/2}(\omega)e^{-i 2m \pi [\lambda-(d-3)/2]}
\,d\lambda,
\end{eqnarray}
which generalizes (\ref{sigma_abs_HR_Poisson}). The steps leading
to (\ref{sigma_abs_SC1}) and (\ref{sigma_abs_SC2}) from
(\ref{sigma_abs_HR_Poisson}) can be trivially repeated here. We
then obtain
\begin{eqnarray}\label{sigma_abs_SC1_ddim}
\sigma_\mathrm{abs}(\omega)&=&\frac{2
\pi^{(d-2)/2}}{\Gamma\left[(d-2)/2\right] \omega^{d-2}}
\int_0^{+\infty} \frac{\Gamma[\lambda+(d-3)/2]}{\Gamma
[\lambda-(d-5)/2]}\,\lambda \Gamma_{\lambda-(d-3)/2}(\omega)
\,d\lambda \nonumber\\
&&+\frac{8 \pi^{d/2}}{\Gamma\left[(d-2)/2\right]
\omega^{d-2}}\,\mathrm{Re} \left( \sum_{m=1}^{+\infty}
\sum_{n=1}^{+\infty} i\,
\frac{\Gamma[\lambda_n(\omega)+(d-3)/2]}{\Gamma
[\lambda_n(\omega)-(d-5)/2]} \lambda_n(\omega)
\gamma_n(\omega)\, e^{i2m \pi[\lambda_n(\omega)-(d-3)/2]} \right)
\nonumber\\
&&+ \frac{2 \pi^{(d-2)/2}}{\Gamma\left[(d-2)/2\right]
\omega^{d-2}} \sum_{m=1}^{+\infty} \left( \int_0^{+i\infty}
\frac{\Gamma[\lambda+(d-3)/2]}{\Gamma [\lambda-(d-5)/2]} \lambda
\Gamma_{\lambda-(d-3)/2}(\omega)\, e^{i2m
\pi[\lambda -(d-3)/2]}\,d\lambda \right. \nonumber\\
&&\hphantom{+ \frac{2 \pi^{(d-2)/2}}{\Gamma\left[(d-2)/2\right]
\omega^{d-2}} \sum_{m=1}^{+\infty} \biggl(}\left. + \int_0^{-i\infty}
\frac{\Gamma[\lambda+(d-3)/2]}{\Gamma [\lambda-(d-5)/2]} \lambda
\Gamma_{\lambda-(d-3)/2}(\omega)\, e^{- i2m \pi[\lambda -(d-3)/2]}
\,d\lambda \right),
\end{eqnarray}
and
\begin{eqnarray}\label{sigma_abs_SC2_ddim}
\sigma_\mathrm{abs}(\omega)&=&\frac{2
\pi^{(d-2)/2}}{\Gamma\left[(d-2)/2\right] \omega^{d-2}}
\int_0^{+\infty} \frac{\Gamma[\lambda+(d-3)/2]}{\Gamma
[\lambda-(d-5)/2]}\,\lambda \Gamma_{\lambda-(d-3)/2}(\omega)
\,d\lambda \nonumber\\
&&- \frac{4 \pi^{d/2}}{\Gamma\left[(d-2)/2\right]
\omega^{d-2}}\,\mathrm{Re} \left( \sum_{n=1}^{+\infty}
\frac{\Gamma[\lambda_n(\omega)+(d-3)/2]}{\Gamma
[\lambda_n(\omega)-(d-5)/2]} \frac{
e^{i\pi[\lambda_n(\omega)-(d-3)/2]}\,\lambda_n(\omega)
\gamma_n(\omega)
}{\sin[\pi (\lambda_n(\omega)-(d-3)/2)]} \right) \nonumber\\
&&+ \frac{\pi^{(d-2)/2}}{\Gamma\left[(d-2)/2\right]
\omega^{d-2}} \int_0^{+i\infty} \left( i\,
\frac{\Gamma[\lambda+(d-3)/2]}{\Gamma [\lambda-(d-5)/2]} \frac{e^{i
\pi[\lambda -(d-3)/2]}\,\lambda
\Gamma_{\lambda-(d-3)/2}(\omega)}{\sin [\pi (\lambda -(d-3)/2)]}
\right. \nonumber\\
&&\hphantom{+ \frac{\pi^{(d-2)/2}}{\Gamma\left[(d-2)/2\right]
\omega^{d-2}} \int_0^{+i\infty} \biggl(}\left. + i\,
\frac{\Gamma[-\lambda+(d-3)/2]}{\Gamma [-\lambda-(d-5)/2]}
\frac{e^{i \pi[\lambda + (d-3)/2]}\,\lambda \Gamma_{-\lambda
-(d-3)/2}(\omega)}{\sin [\pi (\lambda +(d-3)/2)]} \right) d\lambda .
\end{eqnarray}
\end{widetext}

The second term of (\ref{sigma_abs_SC2_ddim}), given by
\begin{eqnarray}\label{Sigma_abs_PR_a_dimD}
&&\sigma_\mathrm{abs}^\mathrm{RP}(\omega)=
-\frac{4\pi^{d/2}}{\Gamma\left[(d-2)/2\right] \omega^{d-2}}\,
\mathrm{Re}
\left( \sum_{n=1}^{+\infty} \right. \nonumber\\
&&\left. 
\frac{\Gamma[\lambda_n(\omega)+(d-3)/2]}
{\Gamma[\lambda_n(\omega)-(d-5)/2]}
\frac{e^{i\pi[\lambda_n(\omega)-(d-3)/2]}
\lambda_n(\omega)\gamma_n(\omega) }{\sin[\pi
(\lambda_n(\omega)-(d-3)/2)]} \right), \nonumber\\
&&
\end{eqnarray}
provides the oscillating part of the total absorption cross
section. In the eikonal regime, (\ref{Sigma_abs_PR_a_dimD})
reduces to a very simple formula. To obtain it, we first note
that (\ref{Modules Tell}) remains valid, but instead of (\ref{V_0
et dersec}) we must now take \cite{DecaniniFolacciRaffaelli2010b}
\begin{subequations} \label{V_0 et dersec_dimD}
\begin{eqnarray}
V_0(\ell)&=&\frac{f(r_c)}{r_c^2}[\ell+(d-3)/2]^2
+ \underset{\ell \to +\infty}{\cal O}(1),\\
V^{''}_0(\ell)&=&-2\left(\frac{\eta_c
f(r_c)}{r_c^2}\right)^2[\ell+(d-3)/2]^2
+ \underset{\ell \to +\infty}{\cal O}(1). \nonumber\\
&&
\end{eqnarray}
\end{subequations}
In the previous equation, the parameter $\eta_c$ is given by
\begin{equation}\label{Eta}
\eta_c=\frac{1}{2}\sqrt{4f(r_c)-2r_{c}^{2}f^{''}(r_c)}.
\end{equation}
It measures the instability of the circular orbits lying on the
photon sphere. Indeed [see Eq.~(28) of
Ref.~\cite{DecaniniFolacciRaffaelli2010b}], it can be expressed in
terms of the Lyapunov exponent $\Lambda_c$ corresponding to these
orbits, introduced in Ref.~\cite{CardosoMirandaBertietal2009}, which
is the inverse of the instability time scale associated with them.
From (\ref{Modules Tell}) and (\ref{V_0 et dersec_dimD}), we obtain
for the residue~(\ref{Sigma_abs_PR_b_ddim})
\begin{equation}\label{Sigma_abs_PR_b_dimD_bis}
\gamma_n(\omega) = -\frac{\eta_c}{2\pi}
+\underset{(r_c/\sqrt{f(r_c)})\omega \to +\infty}{\cal O}\left(
\frac{1}{(r_c/\sqrt{f(r_c)})\omega}\right).
\end{equation}
Furthermore, we recall that in the eikonal regime the Regge poles
$\lambda_{n}(\omega)$ are well described by
\cite{DecaniniFolacciRaffaelli2010b}
\begin{eqnarray}\label{RPHF}
\lambda_{n}(\omega)&=&\frac{r_{c}}{\sqrt{f(r_c)}}~\omega
+i\eta_c \left(n- \frac{1}{2} \right) \nonumber\\
&&+\underset{(r_c/\sqrt{f(r_c)})\omega \to +\infty}{\cal
O}\left( \frac{1}{(r_c/\sqrt{f(r_c)})\omega}\right).\quad
\end{eqnarray}
By inserting (\ref{Sigma_abs_PR_b_dimD_bis}) and (\ref{RPHF})
into (\ref{Sigma_abs_PR_a_dimD}), using (\ref{devS sin_a}) as
well as
\begin{eqnarray} \label{dev_Gamma_sur_Gamma}
&&\frac{\Gamma(z+a)}{\Gamma(z+b)} \sim \left( \frac{1}{z}
\right)^{-a+b} \nonumber\\
&&\qquad\qquad \mathrm{valid\, if}\,\, |z| \to +\infty\,\,
\mathrm{and}\,\, |\arg z| < \pi,
\end{eqnarray}
and keeping only the contribution of the first Regge pole, we
obtain for the high-energy behavior of the oscillating part of
the absorption cross section the very simple formula
\begin{eqnarray}\label{Sigma_abs_PR_a_APP_dimD}
\sigma_\mathrm{abs}^\mathrm{osc}(\omega)&=&(-1)^{d-3}\,
4(d-2) \pi\,\eta_c\, e^{-\pi \eta_c}\,\sigma_\mathrm{geo}
\nonumber\\
&&\times\,\text{sinc}\left[2\pi
(r_c/\sqrt{f(r_c)})\omega\right].
\end{eqnarray}
The constant factor multiplying the sinc function involves again the
geometrical cross section of the black hole
$\sigma_\mathrm{geo}=\pi^{(d-2)/2} b_c^{d-2}/\Gamma(d/2)$ and the
argument of the sinc involves the orbital period $2\pi
(r_c/\sqrt{f(r_c)})=2\pi b_c$ of a massless particle orbiting the
black hole on the photon sphere (see
Ref.~\cite{DecaniniFolacciRaffaelli2010b}). We also note the
presence of a coefficient involving the Lyapunov exponent of the
geodesic followed by the particle, as well as the spacetime
dimension. Formula (\ref{Sigma_abs_PR_a_APP_dimD}) clearly
generalizes (\ref{Sigma_abs_PR_a_APP}), and the universality of the
fluctuations of the high-energy absorption cross sections for black
holes is now obvious.

\begin{figure}
\includegraphics[width=8.5cm]{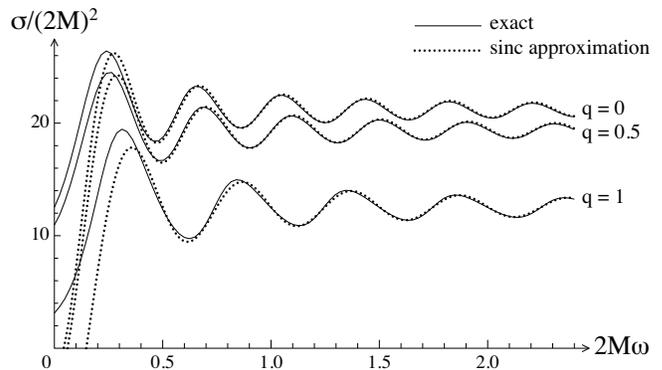}
\caption{\label{fig:RN}Total absorption cross section for a
massless scalar field propagating in the Reissner-Nordstr\"om
geometry for three values of the charge-to-mass ratio $q \equiv
|Q|/M$, as computed in \cite{CrispinoDolanOliveira2009}, compared
to the sum of the geometrical cross section $\sigma_\mathrm{geo}$
with our approximate analytic formula
(\ref{Sigma_abs_PR_a_APP_dimD}). The $q=0$ case was already
illustrated in Fig.~\ref{fig:SecAbsExactAndPR1} above.}
\end{figure}

\begin{figure}
\includegraphics[width=8.5cm]{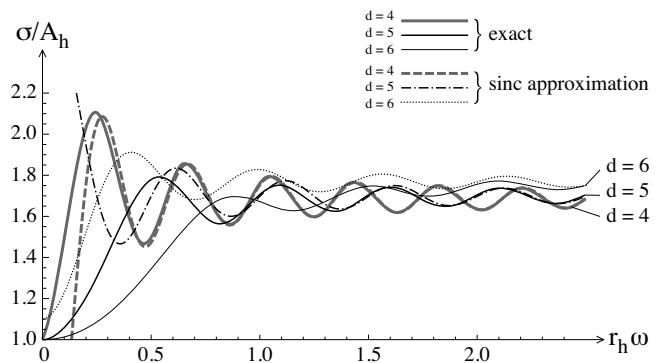}
\caption{\label{fig:ST}Total absorption cross section for a
massless scalar field propagating in the $d$-dimensional
Schwarzschild-Tangherlini geometry, as computed in
\cite{JungPark2005}, compared to the sum of the geometrical cross
section $\sigma_\mathrm{geo}$ with our approximate analytic
formula (\ref{Sigma_abs_PR_a_APP_dimD}). As in the text, $r_h$
denotes the horizon radius in Schwarzschild coordinates, and $A_h
= 2\,\pi^{(d-1)/2}\, r_h^{d-2}/\Gamma[(d-1)/2]$ is the area of
the horizon. The $d=4$ case was already illustrated in
Fig.~\ref{fig:SecAbsExactAndPR1} above. Note that in higher
dimensions, higher frequencies $\omega$ are needed for the sinc
formula (\ref{Sigma_abs_PR_a_APP_dimD}) to give a reasonable
approximation.}
\end{figure}

\begin{figure}
\includegraphics[width=8.5cm]{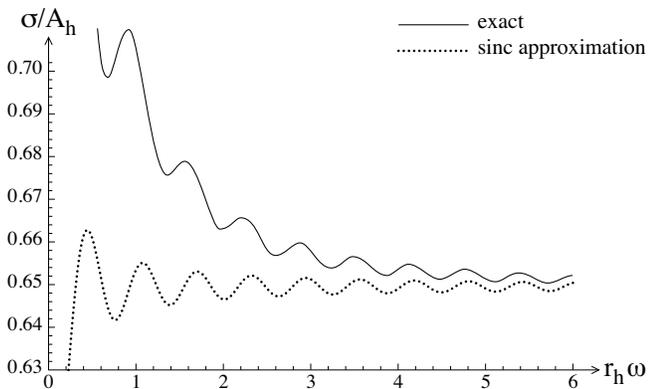}
\caption{\label{fig:AcousticBH}Zoom in on the total absorption
cross section for the canonical acoustic black hole, as computed
in \cite{DolanOliveiraCrispino2009}, compared to the sum of the
geometrical cross section $\sigma_\mathrm{geo}$ with our
approximate analytic formula (\ref{Sigma_abs_PR_a_APP_dimD}). As
before, $A_h = 4 \pi r_h^2$ denotes the horizon area. Note that
in this case of the acoustic black hole, the
$\mathcal{O}(1/\omega^2)$ contribution coming from the first and
third terms of Eq.~(\ref{sigma_abs_SC2_ddim}) is numerically
significant with respect to the oscillations due to the Regge
poles (\ref{Sigma_abs_PR_a_dimD}).}
\end{figure}

We have tested (\ref{Sigma_abs_PR_a_APP_dimD}) by comparing its
predictions with the numerical results displayed in
Refs.~\cite{JungPark2005,CrispinoDolanOliveira2009,
DolanOliveiraCrispino2009}, devoted to Schwarzschild-Tangherlini
and Reissner-Nordstr\"om black holes (both classes in $d \geq
4$), and to the canonical acoustic black hole. As illustrated in
Figs.~\ref{fig:RN}, \ref{fig:ST} and \ref{fig:AcousticBH}, the
agreement is excellent for high enough frequencies. Formula
(\ref{Sigma_abs_PR_a_APP_dimD}) thus allows us to avoid very time
consuming numerical calculations.

Finally, it should be noted that:

--~The first term of (\ref{sigma_abs_SC2_ddim}) provides the
geometrical cross section of the black hole. Indeed, by noting
that on the positive real $\lambda$ axis the greybody factor
$\Gamma_{\lambda-(d-3)/2}(\omega)$ is roughly described by
\begin{equation}\label{GBF_DeWitt_d}
\Gamma_{\lambda-(d-3)/2}(\omega)= \Theta (b_c \omega - \lambda),
\end{equation}
[or can be described by a more accurate expression based on the
WKB approximation (\ref{Modules Tell}) and formulas (\ref{V_0 et
dersec_dimD})] and by using (\ref{dev_Gamma_sur_Gamma}), we
obtain
\begin{eqnarray}\label{First Term_d}
&&\frac{2 \pi^{(d-2)/2}}{\Gamma\left[(d-2)/2\right] \omega^{d-2}}
\int_0^{+\infty} \frac{\Gamma[\lambda+(d-3)/2]}{\Gamma
[\lambda-(d-5)/2]} \nonumber\\
&&\qquad \times \lambda \Gamma_{\lambda-(d-3)/2}(\omega)
\,d\lambda = \frac{\pi^{(d-2)/2} b_c^{d-2}}{\Gamma\left(d/2\right)}
+ \mathcal{O}\left(\frac{1}{\omega^2} \right). \nonumber\\
&&
\end{eqnarray}

--~The third term of (\ref{sigma_abs_SC2_ddim}) vanishes when $d$ is
odd [this is mainly due to the parity of the function
$(\Gamma[\lambda+(d-3)/2]/\Gamma
[\lambda-(d-5)/2])\Gamma_{\lambda-(d-3)/2}(\omega)$] and is
$\mathcal{O}\left(1/\omega^{d-2} \right)$ when $d$ is even.

\section{Conclusions}
\label{sec5} {}From the complex angular momentum approach, we have
been able to extract the physical information encoded in the sum
over all the partial wave contributions defining the absorption
cross section for a massless scalar field propagating in a static
and spherically symmetric black hole of arbitrary dimension endowed
with a photon sphere. We have then emphasized the universal
character of the fluctuations of this absorption cross section at
high energies. In particular, we have shown that the fluctuations
are fully and quite simply described in terms of Regge poles, i.e.,
from the properties (dispersion relation and damping) of the waves
trapped near the photon sphere [see Eqs.~(\ref{Sigma_abs_PR_a}) and
(\ref{Sigma_abs_PR_a_dimD})] and that, in the eikonal regime, they
are described by a very simple formula involving the geometrical
cross section of the black hole and the characteristics (orbital
frequency and Lyapunov exponent) of the null unstable geodesics
lying on the photon sphere [see Eqs.~(\ref{Sigma_abs_PR_a_APP}) and
(\ref{Sigma_abs_PR_a_APP_dimD})]. From a mathematical point of view,
we can note that this universality is a direct consequence of the
following facts: (i) The Regge poles permits us to describe the
properties of the waves trapped near the photon sphere, and (ii) the
residue of the greybody factors taken at the Regge poles are
approximately constant. We believe that our result could be
naturally extended to more general black holes, including rotating
ones, and to more general field theories. We intend, in the near
future, to accomplish some progress in these directions and to
consider more particularly the gravitational wave theory.

It is interesting to recall that the quasinormal mode frequencies of
the black holes are hidden into the terms $1/\sin[\pi
(\lambda_n(\omega)-1/2)]$ of (\ref{Sigma_abs_PR_a}) and $1/\sin[\pi
(\lambda_n(\omega)-(d-3)/2)]$ of (\ref{Sigma_abs_PR_a_dimD}). As a
consequence, by duality between the Regge poles and the complex
frequencies of the weakly damped quasinormal modes of the black hole
(see Refs.~\cite{DecaniniFJ_cam_bh,DecaniniFolacci2010a,
DecaniniFolacciRaffaelli2010b}), the fluctuations of the high-energy
absorption cross section could be also interpreted in terms of
quasinormal modes.

Finally, it is worth pointing out that, {\it mutatis mutandis},
Hawking radiation could be analyzed as a corollary of our previous
results, because the greybody factors are also present in the
emission spectrum of a black hole or, more precisely, because the
particle emission rate can be expressed in terms of the absorption
cross section \cite{Hawking1975}. To simplify these considerations,
let us focus here on the case of the Schwarzschild black hole, but
we could straightforwardly extend them to more general black holes
by using (\ref{Sigma_abs_PR_a_dimD}) or
(\ref{Sigma_abs_PR_a_APP_dimD}). The number of particles emitted by
this black hole per unit time and unit frequency is given
by~\cite{Hawking1975}
\begin{eqnarray}\label{HawkingEmissionSpectrum}
\frac{d^2N}{d\omega dt}(\omega)&=&\frac{1}{2\pi}
\sum_{\ell=0}^{+\infty} \frac{(2\ell + 1)
\Gamma_\ell(\omega)}{\exp(8\pi M\omega)-1} \nonumber\\
&=&\frac{\omega^2}{2\pi^2}\,\frac{\sigma_\mathrm{abs}(\omega)}
{\exp(8\pi M\omega)-1}\,.
\end{eqnarray}
The fluctuations of the absorption cross section induces
fluctuations in the particle emission rate of the Schwarzschild
black hole. They are very attenuated because the Planck factor is
a cutoff for high frequencies, but they can be numerically
observed and they are very well described by combining
(\ref{Sigma_abs_PR_a}) or (\ref{Sigma_abs_PR_a_APP}) with
(\ref{HawkingEmissionSpectrum}). They have therefore a natural
explanation in terms of Regge poles or, equivalently, in terms of
complex quasinormal frequencies.

\bigskip
\begin{acknowledgments}

Y. D. and A. F. thank Bernard Raffaelli for various discussions
concerning Regge poles in black hole physics. A. F. also thanks
Institut d'Astrophysique de Paris and Centre de Physique Th\'eorique
de Marseille for their hospitality.

\end{acknowledgments}

\bibliography{BH_Absorption_Cross_Section}

\end{document}